# Echo-based measurement of the speed of sound

Alivia Berg and Michael Courtney
U.S. Air Force Academy, 2354 Fairchild Drive, USAF Academy, CO 80840
Michael.Courtney@usafa.edu

**Abstract**
The speed of sound in air can be measured by popping a balloon next to a microphone a measured distance away from a large flat wall, digitizing the sound waveform, and measuring the time between the sound of popping and return of the echo to the microphone. The round trip distance divided by the time is the speed of sound.

**Introduction**

Echo technology and time based speed of sound measurements are not new,[1-3] but this technique allows a student lab group or teacher performing a demonstration to make precise speed of sound measurements with a notebook computer, Vernier LabQuest, or other sound digitizer that allows the sound pressure vs. time waveform to be viewed.

**Method**

The method employed here used the sound recording and editing program called Audacity.[4] The program was used to digitize, record, and display the sound waveform of a balloon being popped next to a microphone a carefully measured distance (45.72 m) from a large, flat wall (side of a building) from which the sound echos.  The sample rate was 100000 samples per second.  Three balloons were tested.  The weather was sunny with a dew point of 16.1° C and temperature of 18.3° C.  The elevation at the site was 360 m.  Times of popping and echo return were determined by visual inspection of the sound waveform, zooming in and usng the program cursor as necessary, and then subtracting the pop time from the echo return time to determine the round trip sound travel time.  The round trip distance is twice the distance from the wall.

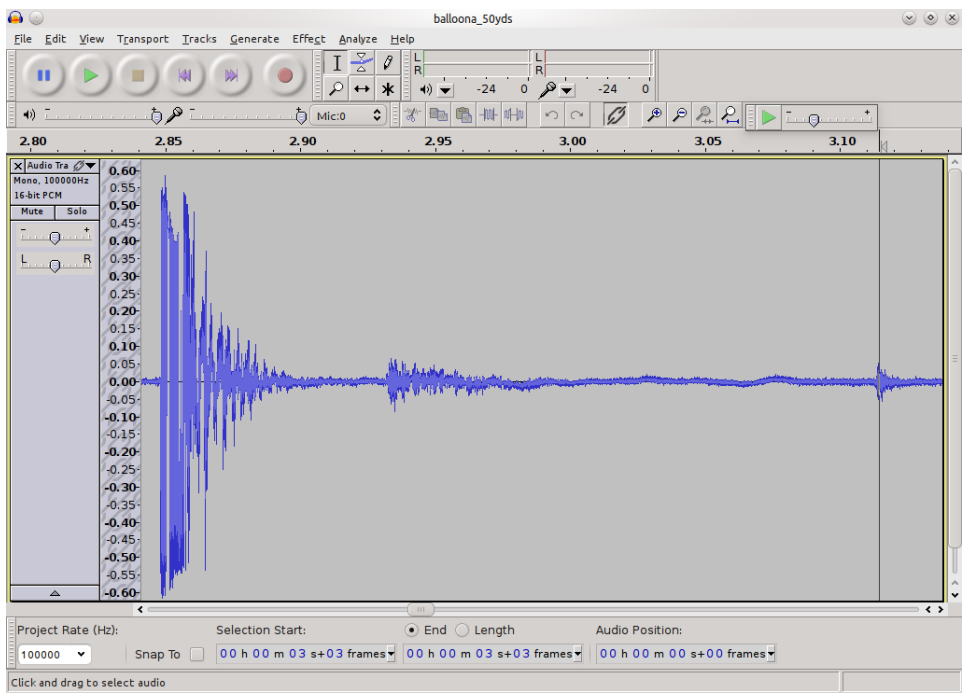

*Figure 1: Balloon A Audacity waveform. Balloon pops near 2.85 s.  The vertical cursor near 3.12s marks the echo.*

**Results**

Figure 1 shows the audio waveform from balloon A at a distance of 45.72 m from the wall (the sound wave had to travel a total of 91.44 m). The balloon popped near 2.85 seconds, and the echo returned near 3.12 seconds. Zooming in and using the cursor to precisely determine the time of popping and echo return yields a round trip travel time of 0.26586 seconds for the sound wave to travel a round trip distance of 91.44 m. The speed of sound is then 343.94 m/s.

Table 1 shows the travel times and resulting speed of sound estimates for three trials which give a mean speed of sound of 344.41(24) m/s, where the digits in parentheses represent the uncertainty of the last two digits. The theoretical (ideal gas) value for speed of sound in dry air at 18.3° C is 342.2 m/s, which is in reasonable agreement with the measured value. Adjusting for the humidity and altitude predicts a speed of sound of 343.4 m/s, (http://resource.npl.co.uk/acoustics/techguides/speedair/) which is in closer agreement and within the uncertainty of the theoretical prediction given the uncertainty of 1.0° C in the temperature measurement.

| Trial | Distance (m) | Time (s) | Speed (m/s) |
|---|---|---|---|
| A | 91.44 | 0.26586 | 343.9404 |
| B | 91.44 | 0.26531 | 344.6534 |
| C | 91.44 | 0.26532 | 344.6404 |

*Table 1: Round trip travel times and resulting speed estimates for sound of balloon popping. Note the round trip distance is twice the distance from the wall, because the popping sound must travel to the wall and return to the microphone.*

**Conclusion**

The echo method for determining travel time is sufficient for determining the speed of sound to better than 0.1%, which is sufficiently accurate to demonstrate the temperature dependence of speed of sound if subsequent measurements are made at times with different ambient temperatures. Another possibility is using the echo technique to determine unknown distances by using the theoretical speed of sound. In either case, this approach puts accurate direct speed of sound measurements within the capabilities of most high-school and undergraduate physics labs who have access to sound digitizing equipment, since Audacity can import and analyze waveforms in a variety of file formats from a variety of sources, including iPods and similar devices.